\documentclass[%
 reprint,
superscriptaddress,
 amsmath,amssymb,
 aps,
 prl,
]{revtex4-1}

\usepackage{graphicx}
\usepackage{dcolumn}
\usepackage{bm}
\usepackage{hyperref}

\usepackage{textgreek}
\usepackage{dsfont}
\usepackage{braket}
\usepackage{xcolor}

\begin{document}

\preprint{APS/123-QED}

\title{Trion photoluminescence and trion stability in atomically thin semiconductors}

\author{Raul Perea-Causin}
\email{causin@chalmers.se}
\affiliation{Department of Physics, Chalmers University of Technology, 412 96 Gothenburg, Sweden}
\author{Samuel Brem}
\affiliation{Department of Physics,
Philipps-Universität Marburg, 35032 Marburg, Germany}
\author{Ole Schmidt}
\affiliation{Department of Physics,
Philipps-Universität Marburg, 35032 Marburg, Germany}
\author{Ermin Malic}
\affiliation{Department of Physics,
Philipps-Universität Marburg, 35032 Marburg, Germany}

\begin{abstract}
The optical response of doped monolayer semiconductors is governed by trions, i.e. photoexcited electron--hole pairs bound to doping charges.
While their photoluminescence (PL) signatures have been identified in experiments, a microscopic model consistently capturing bright and dark trion peaks is still lacking.
In this work, we derive a generalized trion PL formula on a quantum-mechanical footing, considering direct and phonon-assisted recombination mechanisms.
We show the trion energy landscape in WSe\textsubscript{2} by solving the trion Schrödinger equation.
We reveal that the mass imbalance between equal charges results in less stable trions exhibiting a small binding energy and, interestingly, a large energetic offset from exciton peaks in PL spectra. 
Furthermore, we compute the temperature-dependent PL spectra for n- and p-doped monolayers and predict yet unobserved signatures originating from trions with an electron at the {\textLambda} point.
Our work presents an important step towards a microscopic understanding of the internal structure of trions determining their stability and optical fingerprint.
\end{abstract}

\maketitle

Atomically thin semiconductors, such as monolayer transition metal dichalcogenides, exhibit strong Coulomb interactions, providing a platform to investigate quantum many-body physics and showing potential promise for highly-tunable optoelectronic devices\,\cite{wang2018colloquium,mueller2018exciton,perea2022exciton}. The optical response of these materials is dominated by excitons---tightly-bound electron--hole pairs---in the neutral regime\,\cite{he2014tightly,ugeda2014giant,chernikov2014exciton}, and by trions---charged excitons---in the presence of doping\,\cite{mak2013tightly,ross2013electrical}.
The recombination of these charge complexes displays clear spectral signatures in photoluminescence (PL) measurements\,\cite{zhang2015experimental,arora2020dark,selig2016excitonic}.
In particular, dark excitons and trions undergo phonon-assisted recombination\,\cite{brem2020phonon}, exhibiting multiple phonon sidebands in PL spectra\,\cite{he2020valley,liu2019gate,liu2020multipath,yang2022relaxation,rosati2020temporal}.
Specific recombination mechanisms have been identified by measuring g-factors and comparing the energetic offset of the sidebands with the energy of the phonon modes involved.
So far, however, no microscopic model has been put forward to accurately describe PL signatures of \emph{both} bright and dark trions in n- and p-doped TMD monolayers, capturing not only the peak positions, but also their intensity and lineshape on a consistent microcopic footing. This would enable a direct comparison between theoretical predictions and experimental measurements, providing new insights into the recombination channels of different trion species throughout a wide range of experimentally tunable parameters.

\begin{figure}[t!]
    \centering
    \includegraphics[width=\linewidth]{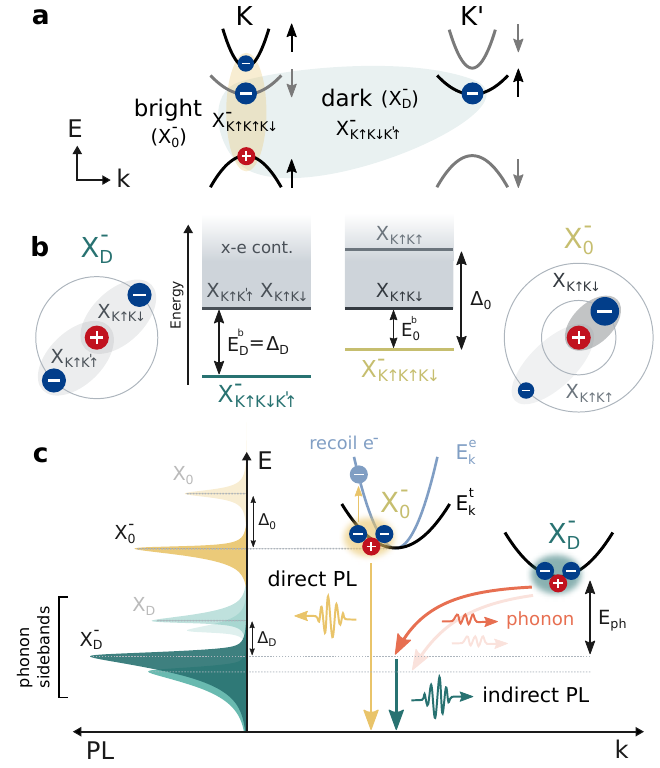}
    \caption{\textbf{a}: Schematic illustration of bright (X$_0^-$) and dark (X$_\text{D}^-$) trions in the band structure of WSe$_2$. The spin-valley configuration of the trion (exciton in b) is denoted in the subscript of X$^-_\text{hee}$ (X$_\text{he}$).
    \textbf{b}: Impact of the mass imbalance on trion stability. Outer panels: visual representation of the electron Bohr radii for equal (X$_\text{D}^-$, $m^\text{e}_\text{K↓}=m^\text{e}_\text{K'↑}$) and unequal (X$_0^-$, $m^\text{e}_\text{K↑} \neq m^\text{e}_\text{K↓}$) electron masses. Central panels: energy of the respective bound trion states and exciton--electron continua (grey shaded).
    The bright (dark) trion binding energy is denoted by $E^\text{b}_{0\text{(D)}}$, and the energy difference with respect to the bright exciton is denoted by $\Delta_{0\text{(D)}}$.
    \textbf{c}: Illustration of bright and dark trion recombination mechanisms. The PL spectrum resulting from the recombination of excitons and trions is sketched on the left.}
    \label{fig1}
\end{figure}

\begin{figure}[t!]
    \centering
    \includegraphics[width=\linewidth]{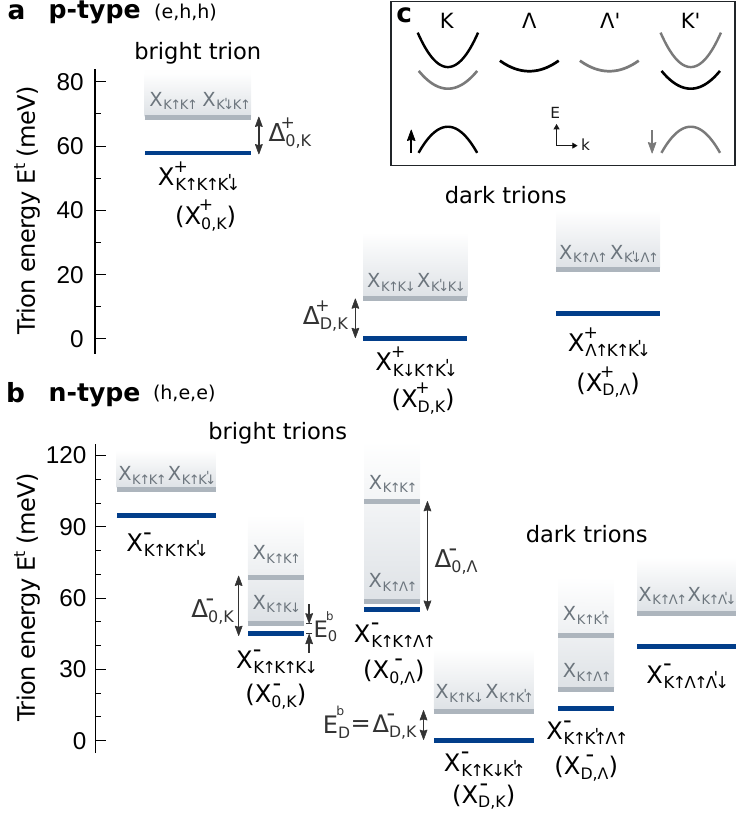}
    \caption{Trion landscape in hBN-encapsulated WSe\textsubscript{2} monolayers (\textbf{a}: p- and \textbf{b}: n-doped).
    Trion and exciton energies are denoted by thick lines. Grey shaded areas illustrate exciton--electron continua. The spin-valley configuration of p- and n-type trions is denoted in the subscript of X$^+_\text{ehh}$ and X$^-_\text{hee}$, respectively. The short notation $\text{X}^\pm_\text{0/D,K/Λ}$ is introduced in the text. Exciton states that can be formed by an electron-hole pair within a trion are denoted by X$_\text{he}$.
    \textbf{c}: Illustration of the electronic band structure with K\textsuperscript{(')} and {\textLambda\textsuperscript{(')}} valleys.}
    \label{fig2}
\end{figure}

In this work, we have developed a microscopic approach to model trion luminescence with a generalized PL formula accounting for direct and phonon-assisted recombination mechanisms (cf. Fig.\,\ref{fig1}). While our approach applies to all atomically thin semiconductors, we consider the exemplary case of tungsten diselenide (WSe\textsubscript{2}), which is known to host both bright and dark trions\,\cite{li2019direct,liu2019gate,liu2020multipath,he2020valley,yang2022relaxation} (cf. Fig.\,\ref{fig1}a).
We gain access to the trion energy landscape via a variational solution of the trion Schrödinger equation, taking into account the multi-valley band structure with an effective mass approximation not only around the K\textsuperscript{(')} point but also around the usually disregarded {\textLambda}\textsuperscript{(')} point (cf. Fig.\,\ref{fig2}c).
Importantly, we incorporate the mass imbalance between the two electrons within a trion (see Fig.\,\ref{fig1}b) in our variational ansatz, leading to less stable trions with a smaller binding energy but, interestingly, to a larger energetic offset from exciton peaks in PL spectra (cf. Fig.\,\ref{fig1}b-c).
Finally, we compute the PL spectra of p- and n-doped monolayers for varying temperatures, obtaining a good agreement with experiments regarding signatures arising from trions formed by K-point electrons and holes. 
Moreover, we predict yet unobserved signatures of trions with an electron located at the {\textLambda} point, which, intriguingly, possess a higher trion energy than their K-point counterpart but exhibit a lower recombination resonance.
We expect that these Λ-point trions play an important role in the relaxation dynamics and they can be experimentally accessed via strain engineering\,\cite{shi2013quasiparticle} or in time-resolved PL spectra\,\cite{rosati2020temporal}.
\\

\noindent \textbf{Microscopic model. }
We consider the multi-valley band structure of the investigated WSe\textsubscript{2} monolayers in an effective-mass approximation around the high-symmetry points K, K', {\textLambda}, and {\textLambda}' (cf. Fig. \ref{fig2}c), with masses and valley offsets from ab initio calculations\,\cite{kormanyos2015k}.
The reversed spin-orbit splitting in this material makes the lowest-energy trion species dark\,\cite{li2019direct,liu2019gate,liu2020multipath,he2020valley,yang2022relaxation} ($\text{X}^-_\text{D}$ in Fig.\,\ref{fig1}a).
Dark trions can still recombine radiatively with the assistance of a phonon providing the necessary momentum or spin (cf. Fig.\,\ref{fig1}c).
In the following, we outline the calculation of the energetic landscape of bright and dark trions and the derivation of the trion PL formula. Further details can be found in the SI.

We obtain the trion eigenenergies $\varepsilon_\nu$ and wave functions $\Psi_\nu(\mathbf{r}_1,\mathbf{r}_2)$ for each spin-valley configuration $\nu=\{\nu_\text{h},\nu_{\text{e}1},\nu_{\text{e}2}\}$ from a variational solution of the Schrödinger equation $\mathcal{H}_\nu \Psi_{\nu}(\mathbf{r}_1,\mathbf{r}_2) = \varepsilon_{\nu} \Psi_{\nu}(\mathbf{r}_1,\mathbf{r}_2)$. Here, $\mathcal{H}_\nu$ is the Hamiltonian for an electron-electron-hole complex\,\cite{berkelbach2013theory} where the interaction between charges is modeled by a thin-film potential
\cite{rytova1967,keldysh1979,laturia2018dielectric}.
The energy $\varepsilon_{\nu}$ is minimized considering the ansatz
\begin{equation}
    \Psi_{\nu}(\mathbf{r}_1,\mathbf{r}_2) \propto \mathrm{e}^{-|\mathbf{r}_1|/a_1}\mathrm{e}^{-|\mathbf{r}_2|/a_2} + C \mathrm{e}^{-|\mathbf{r}_1|/b_1}\mathrm{e}^{-|\mathbf{r}_2|/b_2},
    \label{eq:Psi}
\end{equation}
with $\mathbf{r}_{1(2)}$ being the relative electron--hole coordinate for the electron in the spin--valley state $\nu_\text{e1(2)}$.
This ansatz is similar to the symmetrized product of hydrogenic wave functions\,\cite{berkelbach2013theory,chandrasekhar1944some,lin2022high} but allows for a mass imbalance between the two equal charges within a trion by introducing the weight factor $C$ and by removing the restriction $b_1=a_2,b_2=a_1$. This consideration is highly relevant for n-doped WSe\textsubscript{2}, where the electrons forming a bright trion have different effective masses (Fig.\,\ref{fig1}a).
As our focus lies on the qualitative description of the trion landscape and PL, we disregard the exchange interaction, which is known to lead to small additional splittings of degenerate states\,\cite{yu2014dirac,courtade2017charged}.

To predict the PL spectra arising from the recombination of bright and dark trions, we derive a generalized trion PL formula.
We truncate the Fock space\,\cite{katsch2018theory,perea2022trion} in the limit of low trion density, noting that our model could be extended to higher densities by considering additional many-particle subspaces or the Fermi--polaron picture\,\cite{efimkin2017many,sidler2017fermi,glazov2020optical,tan2020interacting,zipfel2022electron}.
We thus obtain an effective trion Hamiltonian and then exploit Heisenberg's equation with the cluster expansion scheme to obtain a set of coupled equations describing the photon emission rate \cite{brem2020phonon}. Solving these equations in the stationary state yields an expression for the PL intensity arising from the direct and phonon-assisted recombination of trions,
\begin{align}
    &I_\text{PL}(\omega) = \frac{2}{\hbar} \sum_{\nu\nu_\text{e}\mathbf{k}} \frac{\left|\mathcal{M}^{\nu\nu_\text{e}}_\mathbf{k}\right|^2}{(E^\text{t}_{\nu\mathbf{k}}-E^\text{e}_{\nu_\text{e}\mathbf{k}}-\hbar\omega)^2+(\gamma^\text{t}_{\nu\mathbf{k}})^2} \nonumber \\
    & \times \left[ N_{\nu\mathbf{k}} \gamma^\text{t-phot}_{\nu\mathbf{k}} + \sum_{\nu'j\mathbf{q}\pm} \frac{N_{\nu'\mathbf{k}+\mathbf{q}} \left|G^{\nu'\nu}_\mathbf{q}\right|^2 \eta^\pm_{j\mathbf{q}} \gamma^\text{t}_{\nu'\mathbf{k}+\mathbf{q}}}{\left( \Delta E_\mathbf{kq}\right)^2 + \left( \gamma^\text{t}_{\nu'\mathbf{k}+\mathbf{q}} \right)^2} \right] ,
    \label{eq:PL}
\end{align}
with $\Delta E_\mathbf{kq} = E^\text{t}_{\nu'\mathbf{k}+\mathbf{q}} - E^\text{e}_{\nu_\text{e}\mathbf{k}}-\hbar\omega \mp \hbar\Omega_{j\mathbf{q}}$.
The first term in Eq.\,\eqref{eq:PL} describes the direct recombination of a trion with the energy $E^\text{t}_{\nu\mathbf{k}}$ and momentum $\mathbf{k}$ via the emission of a photon and the recoil\,\cite{esser2000photoluminescence,zipfel2022electron} of an electron with energies $\hbar\omega$ and $E^\text{e}_{\nu_\text{e}\mathbf{k}}$, respectively.
The trion--photon matrix element $\mathcal{M}^{\nu\nu_\text{e}}_{\mathbf{k}}$ contains the selection rules $\delta_{\nu_\text{h},\nu_{\text{e}1}}\delta_{\nu_\text{e},\nu_{\text{e}2}}$ or $\delta_{\nu_\text{h},\nu_{\text{e}2}}\delta_{\nu_\text{e},\nu_{\text{e}1}}$ enforcing that the recombining electron and hole are located in the same valley. 
Furthermore, $N_{\nu\mathbf{k}}$ and $\gamma^\text{t}_{\nu\mathbf{k}}$ denote the trion occupation and dephasing, respectively.

The second term in Eq.\,\eqref{eq:PL} accounts for the recombination of a trion in $\nu'$ via the scattering with a phonon with mode $j$, momentum $\mathbf{q}$, and energy $\hbar\Omega_{j\mathbf{q}}$ into a virtual bright state in $\nu$ (cf. Fig.\,\ref{fig1}c).\
The trion in the virtual bright state recombines by emitting a photon and leaving an electron behind.
The trion--phonon matrix element $G^{\nu'\nu}_{j\mathbf{q}}$ is determined by the overlap between initial and final trion states\,\cite{perea2022trion}, together with the carrier-phonon strength which is treated in a deformation potential approach with input parameters obtained from ab initio calculations\,\cite{jin2014intrinsic}.
We have also introduced the factor $\eta^\pm_{j\mathbf{q}} = \rho_\mathbf{q} + (1\pm1)/2$ 
with the phonon number $\rho_\mathbf{q}$.
\\

\noindent \textbf{Trion landscape. }
We now solve the trion Schrödinger equation with the variational ansatz, Eq.\,\eqref{eq:Psi}, for n- and p-doped WSe\textsubscript{2} monolayers encapsulated in hBN. In addition, we compute exciton binding energies with a hydrogenic variational wave function\,\cite{berkelbach2013theory} in order to consistently establish the energetic offset of exciton--electron continua. The calculated trion (and exciton) binding energies are shifted by the respective energy offsets between the single-particle energies of different valleys, yielding the total trion energies shown in Fig.\,\ref{fig2}. 
The spin-valley configuration of p- and n-type trions as well as excitons is denoted by the subindices of X$^+_\text{ehh}$, X$^-_\text{hee}$, and X$_\text{he}$, respectively.
The trion notation for the most relevant states is simplified to $\text{X}^\pm_\text{0/D,K/Λ}$, where the subindices denote a bright/dark (0/D) state and the valley K/Λ of the (additional) electron in p(n)-type trions.
In Fig.\,\ref{fig2} we have categorized the different trion states into bright and dark species according to whether their direct recombination is allowed. Concretely, bright trions have an electron and a hole located at the same valley, occupying bands with equal spin.
In agreement with previous studies\,\cite{deilmann2017dark,liu2019gate,arora2020dark,he2020valley,liu2020multipath,yang2022relaxation}, we find that dark trions formed by electrons and holes at the K\textsuperscript{(')} point, $\text{X}^\pm_\text{D,K}$, constitute the most energetically favourable species for p- and n-type doping (cf. Fig.\,\ref{fig2}).
Dark trions with an electron at the {\textLambda} valley, i.e. $\text{X}^\pm_\text{D,Λ}$, lie only about 10 meV above the lowest state, making them potentially relevant in thermalization, transport, and recombination processes\,\cite{perea2022exciton}.

Interestingly, n-type trions in WSe\textsubscript{2} can be formed by two electrons with a different effective mass due to the variety of conduction-band valleys with masses $m^\text{e}_\text{K↑}=m^\text{e}_\text{K'↓}=0.29\,m_0$, $m^\text{e}_\text{K↓}=m^\text{e}_\text{K'↑}=0.40\,m_0$, and $m^\text{e}_\text{Λ↑}=m^\text{e}_\text{Λ'↓}=0.60\,m_0$\,\cite{kormanyos2015k}. The energy gained by binding either of the two possible excitons to the additional electron can then have two distinct values. The stability of the trion (i.e. trion binding energy) is determined by the lowest of these two (cf. $E^\text{b}$ in Fig.\,\ref{fig1}b), i.e. the lowest energy needed to dissociate the trion into an unbound exciton--electron complex.
In general, we find that trions with mass imbalance are less stable---cf. the smaller binding energy $E^b$ for $\text{X}^-_\text{0,K}$ (mass imbalance) compared to  $\text{X}^-_\text{D,K}$ (mass balance) in Fig.\,\ref{fig2}b.
This is a consequence of the preferential binding of the hole with the heavy electron (cf. Fig.\,\ref{fig1}b). The light electron thus sees a more neutral charge cloud of the other two particles, giving rise to a weak interaction as demonstrated in Fig. 1 in the SI.
Importantly, the optical resonance of trions in PL relative to the exciton is determined by the energy difference between the trion and the bright-exciton--electron state ($\Delta$ in Fig.\,\ref{fig1}b).
The smaller reduced mass and consequent weaker binding of the bright exciton compared to the dark one results in an exciton--trion peak separation larger than the actual trion binding energy, i.e. $\Delta > E^\text{b}$.
In particular, while we predict $\Delta^-_\text{D,K}=12\ \text{meV}$ for the lowest dark species, this value increases to $\Delta^-_\text{0,K}=24\ \text{meV}$ for the lowest n-type bright trion due to its mass imbalance, cf. Fig.\,\ref{fig2}b.
The calculated binding energies for the lowest dark trions (12 meV) closely resemble the experimental measurements (13-14 meV)\,\cite{yang2022relaxation}. The larger exciton--trion energy difference for the n-type bright trion ($\Delta^-_\text{0,K}=24\ \text{meV}$) compared to the p-type ($\Delta^+_\text{0,K}=11\ \text{meV}$) has also been identified experimentally\,\cite{courtade2017charged}, confirming the predictive character of our microscopic theory.

\begin{figure*}[t!]
    \centering
    \includegraphics[width=\linewidth]{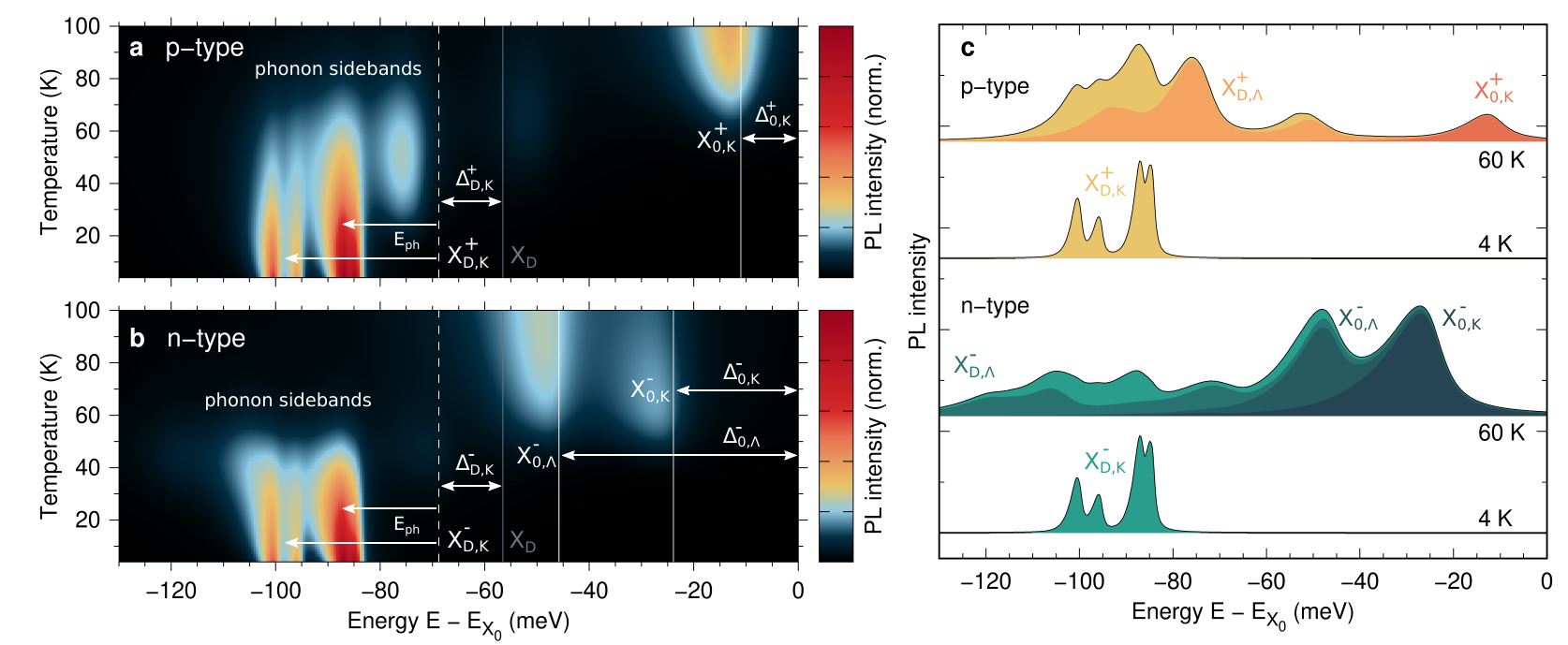}
    \caption{\textbf{a-b}: Temperature-dependent PL spectra in p- and n-type WSe\textsubscript{2}, exhibiting bright ($\text{X}^\pm_\text{0,K/Λ}$, solid white lines) and dark ($\text{X}^\pm_\text{D,K/Λ}$, dashed white lines) trion resonances.
    The origin is offset to the bright exciton $\text{X}_\text{0}=\text{X}_\text{K↑K↑}$ resonance, while the grey line (at approx. -55 meV) denotes the dark exciton $\text{X}_\text{D}=\text{X}_\text{K↑K'↑}$ resonance. The separation between trion and exciton resonances is denoted by $\Delta$. The phonon energy $E_\text{ph}$ determines the position of the phonon sidebands associated with dark trions.
    The spectra have been normalized to the integrated PL for each temperature.
    \textbf{c}: Disentangled contributions to the PL spectra at 4 K and 60 K for p- and n-type doping. Each contribution is labeled with the respective trion state.
    }
    \label{fig3}
\end{figure*}

In addition, our calculations predict that the lowest n-type bright trion, $\text{X}^-_\text{0,K}$, lies only 10 meV below $\text{X}^-_\text{0,Λ}$, with an electron at Λ↑ instead of K↓.
The energetic proximity between these states implies that they can be similarly populated at relatively large temperatures or during high-energy continuous-wave excitation. Interestingly, the larger mass of the Λ electron leads to a larger energy separation from the bright exciton, resulting in an exciton--trion energy difference in PL of $\Delta^-_\text{0,Λ}=46\ \text{meV}$ for $\text{X}^-_\text{0,Λ}$ compared to $\Delta^-_\text{0,K}=24\ \text{meV}$ for $\text{X}^-_\text{0,K}$. Thus, although $\text{X}^-_\text{0,Λ}$ has a higher trion energy $E^\text{t}$, we predict its emission resonance to appear below that of $\text{X}^-_\text{0,K}$.
\\

\noindent \textbf{Trion photoluminescence. }
After computing the trion landscape and understanding the impact of mass imbalance on the trion binding energy, we now investigate the PL signatures arising from the recombination of dark and bright trions. To do so, we evaluate Eq.\,\eqref{eq:PL} for varying temperatures as detailed in the SI.
The obtained temperature maps of the PL spectra for p- and n-doped WSe\textsubscript{2} are shown in Fig.\,\ref{fig3}a-b, with disentangled contributions to the PL at 4 K and 60 K displayed in Fig.\,\ref{fig3}c.
At low temperatures, most of the trion population resides in the dark state $\text{X}^\pm_\text{D,K}$.
This trion species can scatter with a phonon into the virtual bright state $\text{X}^\pm_\text{0,K}$ and then recombine, resulting in a group of peaks (often denoted as phonon sidebands) in PL spectra, each of them shifted from the dark trion resonance by the respective phonon energy.
The PL resonances of dark trions at the lowest temperatures are very similar for p- and n-doping since their binding energy is almost equal.
As the temperature is increased, higher-lying trion states are populated and new PL signatures appear. In particular, at approx. $40\ \text{K}$ the occupation of dark trion states with a {\textLambda} electron results in very distinct PL spectra for p- and n-type doping. On the one hand, $\text{X}^+_\text{D,Λ}$ recombines indirectly via scattering of the Λ↑ electron into K↑, appearing in the spectra as a new peak at $\sim - 75\ \text{meV}$ for p-type doping (Fig.\,\ref{fig3}a and c).
On the other hand, for n-type doping $\text{X}^-_\text{D,Λ}$ can be brightened by scattering of either the K'↑ or Λ↑ electron into K↑, resulting in the low energy tail at $\sim - 105\ \text{meV}$ and the signature at approx. -70 meV, respectively (Fig.\,\ref{fig3}b and c).

At sufficiently large temperatures, bright trions start to dominate the PL spectra, exhibiting resonances marked by white solid lines in Fig.\,\ref{fig3}a-b.
In the case of p-type doping, the only relevant bright trion is $\text{X}^+_\text{0,K}$, which appears in PL 11 meV ($\Delta^+_\text{0,K}$ in Fig.\,\ref{fig3}a) below the bright exciton resonance.
In contrast, n-type samples can host bright trions where the additional electron is at the K ($\text{X}^-_\text{0,K}$) or {\textLambda} ($\text{X}^-_\text{0,Λ}$) valley. As $\text{X}^-_\text{0,K}$ has a lower three-body energy $E^\text{t}$ (cf. Fig.\,\ref{fig2}b), it is populated first and appears in PL at lower temperatures ($\sim 50\ \text{K}$), with a resonance energy $\Delta^-_\text{0,K} = 24\ \text{meV}$ below the bright exciton. Interestingly, $\text{X}^-_\text{0,Λ}$ appears at higher temperatures ($\sim 60\ \text{K}$), but with a lower resonance energy ($\Delta^-_\text{0,Λ} = 46\ \text{meV}$) due to the larger mass of the Λ electron. 
We remark here that, despite the large exciton--trion offset in PL ($\Delta^-_\text{0,K/Λ}$), both n-type bright trions are actually weakly bound (cf. $E^\text{b}_0$ Fig.\,\ref{fig2}b).

The predicted signatures of dark and bright trions composed of charges at the K\textsuperscript{(')} point are in good qualitative agreement with experimental measurements\,\cite{li2019momentum,he2020valley,liu2020multipath,yang2022relaxation}. In particular, our calculations reproduce the binding energy of p- and n-type dark trions and the similar position of their sidebands, as well as the higher energy of the p-type bright trion peak compared to the n-type one.
The predicted PL signatures arising from trions with an electron at the {\textLambda} valley have not been observed in experiments so far, although some peaks still remain unidentified in the literature\,\cite{he2020valley,liu2020multipath}.
The uncertainty in the offset between conduction-band minima at K and {\textLambda}\,\cite{zipfel2020exciton} and the presence of non-equilibrium distributions in experiments might explain why the predicted signatures could not be observed in experimental measurements yet.
Nevertheless, we expect that these trion states, similar to {\textLambda} excitons\,\cite{madeo2020directly,wallauer2021momentum}, play an important role in the thermalization process and could therefore be observed in time-resolved PL experiments\,\cite{rosati2020temporal}.
\\

\noindent \textbf{Conclusion. }
In summary, we have introduced a microscopic approach to describe trion photoluminescence in monolayer semiconductors, accounting for both direct and phonon-assisted recombination processes.
We have considered the exemplary case of WSe\textsubscript{2} and computed the trion energy landscape and the corresponding PL spectra. The qualitative agreement of the PL signatures arising from K-point trions with experiments demonstrates the predictive character of our model, which provides microscopic access into the recombination mechanisms of charge complexes. We have revealed the crucial impact of the electron mass imbalance on the trion stability and trion PL.
Furthermore, we have predicted yet unobserved PL signatures of {\textLambda}-point trions, which we expect to be visible in time-resolved PL experiments. The gained microscopic insights will trigger further experimental and theoretical studies addressing the impact of the {\textLambda} valley on the physics of charge complexes in atomically thin semiconductors.
\\

\begin{acknowledgments}
\noindent \textbf{Acknowledgements. }
We acknowledge funding from the Chalmers' Excellence Initiative Nano under its Excellence PhD program, the Deutsche Forschungsgemeinschaft (DFG) via SFB 1083, and the European Union’s Horizon 2020 research and innovation program under grant agreement no. 881603 (Graphene Flagship).
\end{acknowledgments}


%

\end{document}